# A General Achievable Rate Region for Multiple-Access Relay Channels and Some Certain Capacity Theorems

Mohammad Osmani-Bojd
Department of Electrical Engineering
Ferdowsi University of Mashhad
Mashhad, Iran
mohammad.osmani@gmail.com

Assadallah Sahebalam
Department of Electrical Engineering
Ferdowsi University of Mashhad
Mashhad, Iran
a_sahebalam@yahoo.com

Ghosheh Abed Hodtani
Department of Electrical Engineering
Ferdowsi University of Mashhad
Mashhad, Iran
ghodtani@gmail.com

*Abstract*—In this paper, we obtain a general achievable rate region and some certain capacity theorems for multiple-access relay channel (MARC), using decode and forward (DAF) strategy at the relay, superposition coding at the transmitters. Our general rate region (i) generalizes the achievability part of Slepian-Wolf multiple-access capacity theorem to the MARC, (ii) extends the Cover-El Gamal best achievable rate for the relay channel with DAF strategy to the MARC, (iii) gives the Kramer-Wijengaarden rate region for the MARC, (iv) meets max-flow min-cut upper bound and leads to the capacity regions of some important classes of the MARC.

## I. INTRODUCTION

THE relay channel was first introduced by Van der Meulen [1]. In the Cover-El Gamal seminal paper [2], the relay channel has been studied carefully. In [3], [4], known capacity theorems for the relay channel have been unified into one capacity theorem. Relaying has been proposed as a means to increase coverage area and transmission rate of wireless networks. Relay nodes in cooperation with the users, act as a distributed multi-antenna system. In [5], MARC is introduced, where some sources communicate with one single destination with the help of a relay node. In [6], [7], some capacity regions were determined for the MARC.

### A. Our Motivation and Work

In the literature, we had a general achievable rate for the relay channel [2] and Slepian-Wolf multiple-access capacity theorem [8]; also, extension of the general best rate for the relay channel (Theorem 7, in [2])to a relay network [9], and capacities of special relay channels [2], [10], [11].

In view of the above previous work and motivations, in this paper, we obtain a general achievable rate region for the multiple-access relay channel that might be considered as (i) generalization of Slepian-Wolf multiple-access capacity theorem to the MARC, (ii) extension of the Cover-El Gamal best achievable rate for the relay to the MARC. Also, our motivation, (iii) subsumes the Kramer-Wijengaarden for the MARC and, (iv) meets max-flow min-cut upper bound and leads the capacity region for some important classes of MARC.

### B. Paper Organization

The rest of paper is organized as follow: In section II, we have preliminaries and some definitions . In the section III, we introduce and prove the main theorem. The results of the main theorem are studied in section IV. Finally, we conclude the paper in section V.

## II. PRELIMINARIES

### A. Notation

In this paper, we use the following notations: random variables (r.v.) are denoted by uppercase letters and lowercase letters are used to show their realizations. The probability distribution function (p.d.f) of a r.v. $X$ with alphabet set $\mathcal{X}$ is denoted by $P_X(x)$ where $x \in \mathcal{X}$; $P_{(X|Y)}(x|y)$ denote the conditional p.d.f of $X$ given $Y$, $y \in \mathcal{Y}$. A sequence of r.v.'s $(X_{k,1}, \cdots, X_{k,n})$ with the same alphabet set $\mathcal{X}$ is denoted by $X_k^n$ and its realization is denoted by $(x_{k,1}, \cdots, x_{k,n})$, where $k$ is index for $k^{th}$ sender. The set of all $\epsilon$-typical $n$-sequences $X^n$ with respect to the p.d.f $P_X(x)$, is denoted by $A_\epsilon^n(X)$.

### B. Slepian-Wolf Multiple-Access Channel Capacity Region

Following Slepian and Wolf [8] for the multiple-access channel (MAC), consider a discrete memoryless MAC $(\mathcal{X}_1 \times \mathcal{X}_2, p(y|x_1,x_2), \mathcal{Y})$ with three independent uniformly distributed messages $w_0 \in [1 : 2^{nR_0}]$, $w_1 \in [1 : 2^{nR_1}]$, and $w_2 \in [1 : 2^{nR_2}]$. The first encoder maps each pair $(w_0, w_1)$ into a codeword $x_1^n(w_0, w_1)$ and the second maps each pair $(w_0, w_2)$ into a codeword $x_2^n(w_0, w_2)$. The capacity region for this channel is given by the set of rate triples $(R_0, R_1, R_2)$ such that

$$R_1 \leq I(X_1; Y|X_2, S) \quad (1a)$$
$$R_2 \leq I(X_2; Y|X_1, S) \quad (1b)$$
$$R_1 + R_2 \leq I(X_1, X_2; Y|S) \quad (1c)$$
$$R_0 + R_1 + R_2 \leq I(X_1, X_2; Y) \quad (1d)$$

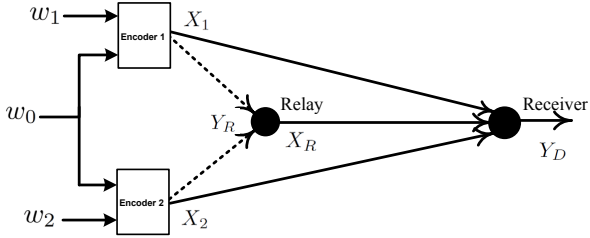

Fig. 1: A two-source multiple-access relay channel

for some $p(s)p(x_1|s)p(x_2|s)$.

## C. Multiple-Access Relay Channel

In multiple-access relay channel, some sources communicate with one single destination with the help of a relay node. An example of such a channel model is the cooperative uplink of some mobile stations to the base station with the help of the relay in a cellular based mobile communication system. Fig. 1 shows a two-source discrete memoryless MARC which is defined by $(\mathcal{X}_1 \times \mathcal{X}_2 \times \mathcal{X}_R, p(y_R, y_D|x_1, x_2, x_R), \mathcal{Y}_R \times \mathcal{Y}_D)$, where $Y_D$ and $Y_R$ are the channel outputs and are received by the receiver and the relay, respectively; $X_k$, $(k = 1, 2)$ and $X_R$ are the channel inputs and are sent by the transmitter and the relay, respectively.

## D. Kramer-Wijengaarden Multiple-Access Relay Channel Achievable Rate Region

In [8], Kramer and Wijengaarden anticipated the following achievable rate region for multiple-access relay channel:

$$R_1 \leq \min\big(I(X_1;Y_R|X_2,X_R), I(X_1,X_R;Y_D|X_2)\big) \quad (2a)$$
$$R_2 \leq \min\big(I(X_2;Y_R|X_1,X_R), I(X_2,X_R;Y_D|X_1)\big) \quad (2b)$$
$$R_1 + R_2 \leq \min\big(I(X_1,X_2;Y_R|X_R), I(X_1,X_2,X_R;Y_D)\big) \quad (2c)$$

where

$$p(x_1, x_2, x_R) = p(x_1)p(x_2)p(x_R|x_1, x_2) \quad (3)$$

## E. Special Classes of Multiple-Access Relay Channel

*1) Multiple-Access Degraded Relay Channel (MADRC):* In multiple-access degraded relay channel, all channels between senders and relay are better than direct channels, such that for MADRC, we have:

$$p(x_1, x_2, x_R, y_R, y_D) = $$
$$p(x_1, x_2, x_R)p(y_D|x_R, y_R)p(y_R|x_1, x_2, x_R) \quad (4)$$

*2) Multiple-Access Reversely Degraded Relay Channel (MARDRC):* In multiple-access reversely degraded relay channel, all channels between senders and receiver are better than channels between senders and relay, hence, we have:

$$p(x_1, x_2, x_R, y_R, y_D) = $$
$$p(x_1, x_2, x_R)p(y_D|x_1, x_2, x_R)p(y_R|y_D, x_R) \quad (5)$$

*3) Multiple-Access Semi-Deterministc Relay Channel (MASDRC):* If $Y_R = g(X_1, X_2, X_R)$ and the senders know the each other messages, then $Y_R$ is known at the senders (assuming that the senders know the first symbol of $X_R$ ), hence, we have:

$$p(x_1, x_2, x_R, y_R, y_D) = $$
$$p(x_1, x_2, x_R, y_R)p(y_D|x_1, x_2, x_R, y_R) \quad (6)$$

*4) Multiple-Access Relay Channel with Orthogonal Components (MARCO):* $X_k$ $(k = 1, 2)$ is divided to orthogonal components $(X_{Rk}, X_{Dk})$ and these components are sent from the senders to the relay $(X_{Rk})$ and from the senders and relay to the receiver $(X_{Dk}, X_R)$. A discrete memoryless multiple-access relay channel is said to have orthogonal components if the channel input-output distribution can be expressed as

$$P(y_D, y_R, x_{R1}, x_{R2}, x_{D1}, x_{D2}, x_R) = P(y_R|x_{R1}, x_{R2}, x_R)$$
$$P(y_D|x_{D1}, x_{D2}, x_R)P(x_R)\prod_{k=1}^{2} P(x_{Rk}|x_R)P(x_{Dk}|x_R) \quad (7)$$

## III. MAIN THEOREM

**Theorem**. A general achievable rate region for two-source multiple-access relay channel is given by $\bigcup\{(R_0, R_1, R_2) :$

$$R_1 \leq \min\Big(I(U_1, X_1; Y_D|U_0, U_2, X_2, X_R) + I(X_R; Y_D), \quad (8a)$$
$$I(U_1; Y_R|U_0, U_2, X_R) + I(X_1; Y_D|U_0, U_1, U_2, X_2, X_R)\Big)$$

$$R_2 \leq \min\Big(I(U_2, X_2; Y_D|U_0, U_1, X_1, X_R) + I(X_R; Y_D), \quad (8b)$$
$$I(U_2; Y_R|U_0, U_1, X_R) + I(X_2; Y_D|U_0, U_1, U_2, X_1, X_R)\Big)$$

$$R_1 + R_2 \leq \min\Big(I(U_1, U_2, X_1, X_2; Y_D|U_0, X_R) + I(X_R; Y_D),$$
$$I(U_1, U_2; Y_R|U_0, X_R) + I(X_1, X_2; Y_D|U_0, U_1, U_2, X_R)\Big) \quad (8c)$$

$$R_0 + R_1 + R_2 \leq \min\Big((I(X_1, X_2, X_R; Y_D), \quad (8d)$$
$$I(U_0, U_1, U_2; Y_R|X_R) + I(X_1, X_2; Y_D|U_0, U_1, U_2, X_R)\Big)\}$$

where the union is taken over all $p(x_1, x_2, x_R, u_0, u_1, u_2)$ for which

$$p(x_1, x_2, x_R, u_0, u_1, u_2) = $$
$$P(x_R)P(u_0|x_R)\prod_{k=1}^{2} P(u_k|u_0, x_R)P(x_k|u_0, u_k, x_R) \quad (9)$$

### A. Proof of the Theorem

We split every message $w_k$ into two parts $w'_k$ and $w''_k$ with respective rates $R'_k$ and $R''_k$. We consider $B$ blocks, each of $n$ symbols. We use superposition coding. In each block, $b = 1, 2, \cdots, B+1$, we use the same set of codebooks:

$$\mathcal{C} = \{x_R^n(m), u_0^n(j, m), u_1(j, m, l_1), u_2(j, m, l_2),$$
$$x_1(j, m, l_1, q_1), x_2(j, m, l_2, q_2)\}$$
$$m = (m_1, m_2) \in [1 : 2^{nR'_1}] \times [1 : 2^{nR'_2}] \quad R = R'_1 + R'_2,$$
$$j \in [1 : 2^{nR_0}], l_k \in [1 : 2^{nR'_k}], \quad q_k \in [1 : 2^{nR''_k}], k = 1, 2.$$

Now, we proceed with proof of achievability using a random coding technique.

*Random codebook generation:* First, fix a choice of $P(u_0, u_1, u_2, x_R, x_1, x_2) = P(x_R)P(u_0|x_R)\prod_{k=1}^{2} P(u_k|u_0, x_R)P(x_k|u_0, u_k, x_R)$

1) Generate $2^{nR}$ independent identically distributed $n$-sequence $x_R^n$, each drawn according to $P(x_R^n) = \prod_{t=1}^{n} P(x_{R,t})$ and index them as $x_R^n(m)$, $m \in [1:2^{nR}]$.
2) For each $x_R^n(m)$ generate $2^{nR_0}$ conditionally independent $n$-sequence $u_0^n$, each drawn according to $P(u_0^n|x_R^n) = \prod_{t=1}^{n} P(u_{0,t}|x_{R,t})$. Index them as $u_0^n(j,m)$, $j \in [1:2^{nR_0}]$.
3) For each $\{x_R^n(m), u_0^n(j,m)\}$, generate $2^{nR_k'}$, $k=1,2$, conditionally independent $n$-sequence $u_k^n$, each drawn according to $P(u_k^n|x_R^n(m), u_0^n(j,m)) = \prod_{t=1}^{n} P(u_{k,t}|x_{R,t}(m), u_{0,t}(j,m))$. Index them as $u_k^n(j,m,l_k)$, $l_k \in [1:2^{nR_k'}]$.
4) For each $\{x_R^n(m), u_0^n(j,m), u_k^n(j,m,l_k)\}$, generate $2^{nR_k''}$, $k=1,2$, conditionally independent $n$-sequence $x_k^n$, each drawn according to $P(x_k^n|x_R^n(m), u_0^n(j,m), u_k^n(j,m,l_k)) = \prod_{t=1}^{n} P(x_{k,t}|x_{R,t}(m), u_{0,t}(j,m), u_{k,t}(j,m,l_k))$. Index them as $x_k^n(j,m,l_k,q_k)$, $q_k \in [1:2^{nR_k''}]$.
5) Partition the sequence $(u_k^n, x_k^n)$, $k=1,2$, into $2^{nR}$ bins, randomly.
6) Partition the sequence $(u_1^n, x_1^n, u_2^n, x_2^n)$ into $2^{nR}$ bins, randomly.
7) Partition the sequence $(u_0^n, u_1^n, x_1^n, u_2^n, x_2^n)$ into $2^{nR}$ bins, randomly.

*Encoding:* Encoding is performed in $B+1$ blocks. The encoding strategy is shown in table I.

1) *Source Terminals:* The message $w_0$ and $w_k'$ are split into $B$ equally sized blocks $w_{0,b}, w_{k,b}'$, $k=1,2$, $b=1,\ldots,B$. Similarly, $w_k''$ is split into $B$ equally sized blocks $w_{k,b}''$, $k=1,2$, $b=1,\ldots,B$. In block $b=1,\cdots,B+1$, the $k^{th}$ encoder sends $x_{k,b}^n(w_{0,b}, w_{k,b-1}', w_{k,b}', w_{k,b}'')$ over the channel.

2) *Relay Terminal:* After the transmission of block $b$ is completed, the relay has seen $y_{R,b}^n$. The relay tries to find $\widetilde{w}_{0,b}, \widetilde{w}_{1,b}'$ and $\widetilde{w}_{2,b}'$ such that

$$\Big(u_{1,b}^n(\widetilde{w}_{0,b}, \hat{w}_{1,b-1}', \widetilde{w}_{1,b}'), u_{2,b}^n(\widetilde{w}_{0,b}, \hat{w}_{2,b-1}', \widetilde{w}_{2,b}'),$$
$$x_{R,b}^n(\hat{w}_{1,b-1}', \hat{w}_{2,b-1}'), u_{0,b}^n(\hat{w}_{0,b-1}, \hat{w}_{1,b-1}', \hat{w}_{2,b-1}'),$$
$$y_{R,b}^n\Big) \in A_\epsilon^n(U_1, U_2, X_R, U_0, Y_R) \quad (10)$$

where $\hat{w}_{0,b-1}, \hat{w}_{1,b-1}'$ and $\hat{w}_{2,b-1}'$ are the relay terminal's estimate of $w_{0,b-1}, w_{1,b-1}'$ and $w_{2,b-1}'$, respectively. If one or more such $w_{1,b}'$ and $w_{2,b}'$ are found, then the relay chooses one of them, and then transmits $x_{R,b+1}^n(\hat{w}_{1,b}', \hat{w}_{2,b}')$ in block $b+1$.

3) *Sink Terminal:* After block $b$, the receiver has seen $y_{D,b-1}^n$ and $y_{D,b}^n$ and tries to find $\widetilde{w}_{0,b-1}$, $\widetilde{w}_{1,b-1}', \widetilde{w}_{2,b-1}', \widetilde{w}_{1,b-1}''$ and $\widetilde{w}_{2,b-1}''$ such that

$$\Big(x_{R,b}^n(\widetilde{w}_{1,b-1}', \widetilde{w}_{2,b-1}'), y_{D,b}^n\Big) \in A_\epsilon^n(X_R, Y_D) \quad (11)$$

and
$$\Big(u_{1,b-1}^n(\widetilde{w}_{0,b-1}, \hat{w}_{1,b-2}', \widetilde{w}_{1,b-1}'),$$
$$u_{2,b-1}^n(\widetilde{w}_{0,b-1}, \hat{w}_{2,b-1}', \widetilde{w}_{2,b-1}'),$$
$$u_{0,b-1}^n(\hat{w}_{0,b-2}, \hat{w}_{1,b-2}', \hat{w}_{2,b-2}'),$$
$$x_{1,b-1}^n(\widetilde{w}_{0,b-1}, \hat{w}_{1,b-2}', \widetilde{w}_{1,b-1}', \widetilde{w}_{1,b-1}''),$$
$$x_{2,b-1}^n(\widetilde{w}_{0,b-1}, \hat{w}_{2,b-2}', \widetilde{w}_{2,b-1}', \widetilde{w}_{2,b-1}''),$$
$$x_{R,b-1}^n(\hat{w}_{1,b-2}', \hat{w}_{2,b-2}'), y_{D,b-1}^n\Big)$$
$$\in A_\epsilon^n(U_1, U_2, U_0, X_1, X_2, X_R, Y_D) \quad (12)$$

*Decoding and error Analysis:* It can be shown that the relay, after determining $x_R^n$ from $y_R^n$, uses jointly decoding and can decode reliably if

$$R_1' \leq I(U_1; Y_R|X_R, U_2, U_0) \quad (13)$$
$$R_2' \leq I(U_2; Y_R|X_R, U_1, U_0) \quad (14)$$
$$R_1' + R_2' \leq I(U_1, U_2; Y_R|X_R, U_0) \quad (15)$$
$$R_0 + R_1' + R_2' \leq I(U_0, U_1, U_2; Y_R|X_R) \quad (16)$$

and the receiver decodes $x_R^n$, $u_0^n$ and other messages with arbitrarily small probability of error if

$$R_1'' \leq I(X_1; Y_D|U_0, U_1, U_2, X_2, X_R) \quad (17)$$
$$R_2'' \leq I(X_2; Y_D|U_0, U_1, U_2, X_1, X_R) \quad (18)$$
$$R_1'' + R_2'' \leq I(X_1, X_2; Y_D|U_0, U_1, U_2, X_R) \quad (19)$$
$$R_1' + R_1'' - I(X_R; Y_D)$$
$$\leq I(U_1, X_1; Y_D|U_2, U_0, X_2, X_R) \quad (20)$$
$$R_2' + R_2'' - I(X_R; Y_D)$$
$$\leq I(U_2, X_2; Y_D|U_1, U_0, X_1, X_R) \quad (21)$$
$$R_1' + R_1'' + R_2' + R_2'' - I(X_R; Y_D)$$
$$\leq I(U_1, U_2, X_1, X_2; Y_D|U_0, X_R) \quad (22)$$
$$R_0 + R_1' + R_1'' + R_2' + R_2'' - I(X_R; Y_D)$$
$$\leq I(U_0, U_1, U_2, X_1, X_2; Y_D|X_R) \quad (23)$$

Therefore, by fully considering (13)-(23), the theorem is proved.

## IV. THE RESULTS OF THE THEOREM

For the multiple-access relay channel, the achievable rate regions and for some special cases, the capacity regions has been found in the following:

### A. Achievability Region of Slepian-Wolf Multiple-Access Channel

Suppose that $X_k$, $k=1,2$, sees a source of rate $R_k$, and in addition, all $X_k$ see a common source of rate $R_0$. All three sources are independent. To obtain the achievable rate region let $U_0 = S$ and $U_k = X_k$. With removing the relay, we obtain achievability of $(R_0, R_1, R_2)$ according to the (1a)-(1d).

### B. The achievable Rate Region of Kramer-Wijengaarden Work

The anticipated Kramer-Wijengaarden multiple-access relay channel achievable rate region in accordance with our definition, is multiple-access degraded relay channel which is discussed in the next section (C-1).

TABLE I: Encoding Strategy

| Block1 | Block2 | $\cdots$ | Block B+1 |
|---|---|---|---|
| $u_{0,1}^n(1,1,1)$ | $u_{0,2}^n(w_{0,1},w'_{1,1},w'_{2,1})$ | $\cdots$ | $u_{0,B+1}^n(w_{0,B},w'_{1,B},w'_{2,B})$ |
| $x_{R,1}^n(1,1)$ | $x_{R,2}^n(w'_{1,1},w'_{2,1})$ | $\cdots$ | $x_{R,B+1}^n(w'_{1,B},w'_{2,B})$ |
| $u_{1,1}^n(w_{0,1},1,w'_{1,1})$ | $u_{1,2}^n(w_{0,2},w'_{1,1},w'_{1,2})$ | $\cdots$ | $u_{1,B+1}^n(1,w'_{1,B},1)$ |
| $u_{2,1}^n(w_{0,1},1,w'_{2,1})$ | $u_{2,2}^n(w_{0,2},w'_{2,1},w'_{2,2})$ | $\cdots$ | $u_{2,B+1}^n(1,w'_{2,B},1)$ |
| $x_{1,1}^n(w_{0,1},1,w'_{1,1},w''_{1,1})$ | $x_{1,2}^n(w_{0,2},w'_{1,1},w'_{1,2},w''_{1,2})$ | $\cdots$ | $x_{1,B+1}^n(1,w'_{1,B},1,1)$ |
| $x_{2,1}^n(w_{0,1},1,w'_{2,1},w''_{2,1})$ | $x_{2,2}^n(w_{0,2},w'_{2,1},w'_{2,2},w''_{2,2})$ | $\cdots$ | $x_{2,B+1}^n(1,w'_{2,B},1,1)$ |

*C. Capacity Region of Some Special Classes of Multiple-Access Relay Channel*

*1) The Capacity Region of Multiple-Access Degraded Relay Channel :* With substitution $U_k = X_k$ or $U_k = f_k(X_k)$ ($f_k$ is reversible), $k = 1, 2$, and $U_0 = \phi$ in (8a)-(8d) then MARC is an MADRC ($(X_1, X_2) \to (X_R, Y_R) \to Y_D$) and there exists $p(x_1, x_2, x_R, u_0, u_1, u_2) = p(x_1, x_2, x_R)$ such that for MADRC, we have:

$$p(x_1, x_2, x_R, y_R, y_D) = \qquad (24)$$
$$p(x_1, x_2, x_R)p(y_D|x_R, y_R)p(y_R|x_1, x_2, x_R)$$

$$R_1 \leq \min\big(I(X_1; Y_R|X_2, X_R), I(X_1, X_R; Y_D|X_2)\big) \quad (25a)$$
$$R_2 \leq \min\big(I(X_2; Y_R|X_1, X_R), I(X_2, X_R; Y_D|X_1)\big) \quad (25b)$$
$$R_1 + R_2 \leq \min\big(I(X_1, X_2; Y_R|X_R), I(X_1, X_2, X_R; Y_D)\big) \quad (25c)$$

It is shown in [7] that achievable rate in (25a)-(25c) meets its outer bound; therefore, the above achievable rate is a capacity rate region.

*2) The Capacity Region of Multiple-Access Reversely Degraded Relay Channel :* In multiple-access reversely degraded relay channel $U_k = X_R$, $k = 1, 2$, and $U_0 = \phi$. We have $(X_1, X_2) \to (X_R, Y_D) \to Y_R$. Consequently, MARC is an MARDRC and there exists $p(x_1, x_2, x_R, u_0, u_1, u_2) = p(x_1, x_2, x_R)$ such that for MARDRC, we have:

$$p(x_1, x_2, x_R, y_R, y_D) = \qquad (26)$$
$$p(x_1, x_2, x_R)p(y_D|x_1, x_2, x_R)p(y_R|y_D, x_R)$$

It is easy to show that the capacity region for MARDRC is as following

$$R_1 \leq I(X_1; Y_D|X_2, X_R) \quad (27a)$$
$$R_2 \leq I(X_2; Y_D|, X_1, X_R) \quad (27b)$$
$$R_1 + R_2 \leq I(X_1, X_2; Y_D|X_R) \quad (27c)$$

*3) The Capacity Region of Multiple-Access Semi-Deterministic Relay Channel :* If $Y_R = g(X_1, X_2, X_R)$ and the senders know the each other messages, then $Y_R$ is known at the senders and according to the lemma in [3] and [4], $U_k$, $k = 1, 2$, is also a function of $X_k$, $k = 1, 2$, and $X_R$ and we have $U_k = Y_R$, then there exists, $p(x_1, x_2, x_R, u_0, u_1, u_2) = p(x_1, x_2, x_R, y_R)$ such that for MASDRC, we have:

$$p(x_1, x_2, x_R, y_R, y_D) = p(x_1, x_2, x_R, y_R)p(y_D|x_1, x_2, x_R, y_R)$$

We obtain achievability of $(R_1, R_2)$ as following,

$$R_1 \leq I(X_1; Y_D|Y_R, X_2, X_R) \quad (28a)$$
$$R_2 \leq I(X_2; Y_D|Y_R, X_1, X_R) \quad (28b)$$
$$R_1 + R_2 \leq \min\big(H(Y_R|X_R) + I(X_1, X_2; Y_D|Y_R, X_R) \\ I(Y_R, X_1, X_2; Y_D|X_R) + I(X_R; Y_D)\big) \quad (28c)$$

It is easy to show that this rate is also an outer bound for MASDERC.

*4) The Capacity Region of Multiple-Access Relay Channel with Orthogonal Components:* In [6], a capacity region of MARCO was obtained. If $X_k = (X_{Rk}, X_{Dk})$, $k = 1, 2$, and $U_k = X_{Rk}$, $k = 1, 2$; therefore, a capacity region is obtained as following:

$$R_1 \leq \min\big(I(X_{D1}, X_R; Y_D|X_{D2}), \quad (29a)$$
$$I(X_{R1}; Y_R|X_{R2}, X_R) + I(X_{D1}; Y_D|X_{D2}, X_R)\big)$$
$$R_2 \leq \min\big(I(X_{D2}, X_R; Y_D|X_{D1}), \quad (29b)$$
$$I(X_{R2}; Y_R|X_{R1}, X_R) + I(X_{D2}; Y_D|X_{D1}, X_R)\big)$$
$$R_1 + R_2 \leq \min\big(I(X_{D1}, X_{D2}, X_R; Y_D), \quad (29c)$$
$$I(X_{R1}, X_{R2}; Y_R|X_R) + I(X_{D1}, X_{D2}; Y_D|X_R)\big)$$

where

$$P(x_{R1}, x_{R2}, x_{D1}, x_{D2}, x_R)$$
$$= P(x_R) \prod_{k=1}^{2} P(x_{Rk}|x_R)P(x_{Dk}|x_R) \quad (30)$$

V. CONCLUSION

We obtain a general achievable rate region and some certain capacity theorems for MARC. Our general rate region generalizes the achievability part of Slepian-Wolf multiple-access capacity theorem to the MARC, extends the Cover-El Gamal best achievable rate for the relay channel with DAF strategy to the MARC, gives the Kramer-Wijengaarden rate region for the MARC, meets max-flow min-cut upper bound and leads to the capacity regions of some important classes of the MARC such as MADRC, MARDRC, MASDRC and MARCO.